\documentclass[conference]{IEEEtran}
\IEEEoverridecommandlockouts
\usepackage{cite}
\usepackage{amsmath,amssymb,amsfonts}
\usepackage{algorithmic}
\usepackage{graphicx}
\usepackage{textcomp}
\usepackage{xcolor}
\usepackage{booktabs}
\def\BibTeX{{\rm B\kern-.05em{\sc i\kern-.025em b}\kern-.08em
    T\kern-.1667em\lower.7ex\hbox{E}\kern-.125emX}}
\DeclareRobustCommand*{\IEEEauthorrefmark}[1]{%
    \raisebox{0pt}[0pt][0pt]{\textsuperscript{\footnotesize\ensuremath{#1}}}}
\renewcommand{\footnoterule}{%
  \kern -3pt
  \hrule width 0.2\textwidth height 0.5pt
  \kern 2pt
}
\begin{document}

\title{Stepwise Schema-Guided Prompting Framework with Parameter Efficient Instruction Tuning for Multimedia Event Extraction
\thanks{\textcopyright\ 2025 IEEE. Personal use of this material is permitted. Permission from IEEE must be obtained for all other uses, in any current or future media, including reprinting/republishing this material for advertising or promotional purposes, creating new collective works, for resale or redistribution to servers or lists, or reuse of any copyrighted component of this work in other works. DOI: 10.1109/ICME59968.2025.11210082. This work is supported by the National Key R\&D Program of China (2023YFC3304902).}
}

\author{
  \IEEEauthorblockN{
    Xiang Yuan\IEEEauthorrefmark{1}\textsuperscript{*$\dagger$},
    Xinrong Chen\IEEEauthorrefmark{1}\textsuperscript{*},
    Haochen Li\IEEEauthorrefmark{2},
    Hang Yang\IEEEauthorrefmark{3},
    Guanyu Wang\IEEEauthorrefmark{1},
    Weiping Li\IEEEauthorrefmark{1},
    Tong Mo\IEEEauthorrefmark{1}\textsuperscript{$\ddag$}
  }
  \IEEEauthorblockA{\IEEEauthorrefmark{1} School of Software and Microelectronics, Peking University, Beijing, China}
  \IEEEauthorblockA{\IEEEauthorrefmark{2} 01.AI, Beijing, China \quad \IEEEauthorrefmark{3} Baidu Inc., Beijing, China}
  \IEEEauthorblockA{xiangyuan@stu.pku.edu.cn, motong@ss.pku.edu.cn}
}

\maketitle
\begingroup\renewcommand\thefootnote{*}\footnotetext[1]{Xiang Yuan and Xinrong Chen contributed equally to this work.} 
\renewcommand\thefootnote{$\dagger$}\footnotetext[2]{Part of this work was done during Xiang Yuan's internship at Baidu Inc.} 
\renewcommand\thefootnote{$\ddag$}\footnotetext[3]{Tong Mo is the corresponding author (motong@ss.pku.edu.cn).}
\endgroup

\maketitle

\renewcommand{\thefootnote}{}

\begin{abstract}
Multimedia Event Extraction (MEE) has become an important task in information extraction research as news today increasingly prefers to contain multimedia content. Current MEE works mainly face two challenges: (1) Inadequate extraction framework modeling for handling complex and flexible multimedia event structure; (2) The absence of multimodal-aligned training data for effective knowledge transfer to MEE task. In this work, we propose a Stepwise Schema-Guided Prompting Framework (SSGPF) using Multimodal Large Language Model (MLLM) as backbone for adaptive structure capturing to solve MEE task. At the initial step of SSGPF, we design Event Type Schema Guided Prompting (ETSGP) for event detection, then we devise Argument Role Schema Guided Prompting (ARSGP) that contains multi-step prompts with text-bridged grounding technique for argument extraction. We construct a weakly-aligned multimodal event labeled dataset based on existing unimodal event annotations, then conduct parameter efficient instruction tuning with LoRA on LLaVA-v1.5-7B under SSGPF. Experiments on the M2E2 benchmark demonstrate that SSGPF significantly outperforms current SOTA baselines by 5.8 percent F1 on event detection and 8.4 percent F1 on argument extraction.
\end{abstract}


\section{Introduction}
\label{sec:intro}
\begin{figure}[t]
\centerline{\includegraphics{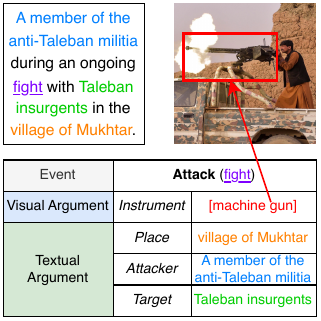}}
\caption{Example of Multimedia Event Extraction. Given the multimodal input, model needs to extract event type, trigger word, and spot arguments belonging to each role label predefined in corresponding event type's argument schema.}
\label{fig_intro}
\vspace{-1em}
\end{figure}

With the rapid development of social media platforms, news on the Internet today tends to combine both textual and visual information. Compared with traditional text-only news recording, multimodal news provides a more accurate and comprehensive description of the event, which can resolve ambiguity caused by unimodal input and enable crossmodal information supplementation. This has led to the emergence of Multimedia Event Extraction (MEE) task in the field of Information Extraction (IE) research, which aims to extract Multimedia Event Structure (MES) from the given news-related text and image. MEE contains two subtasks: Multimedia Event Detection (MED) and Multimedia Event Argument Extraction (MEAE). As shown in Fig. \ref{fig_intro}, MED aims to identify the event type \textit{Attack} depicted by the multimodal input and the trigger word \textit{fight} in text expressing such event type, then MEAE intends to extract arguments in each modality that constitute the event with role labels predefined in event schema, such as \textit{Taleban insurgents} in textual modality with role label \textit{Target} and the bounding box region that describes the object ``machine gun'' in visual modality representing role label \textit{Instrument}. The MEE results reveal the \textbf{complexity and flexibility} of MES, as it encompasses various structured information with arguments derived from multiple modalities, and different event types contain distinct candidate argument role labels predefined by flexible event schema.

Current works mainly face two challenges when addressing MEE task. The first is the \textbf{inadequate extraction framework modeling for handling complex and flexible MES}. WASE \cite{wase}, UniCL \cite{unicl}, CAMEL \cite{camel}, and MGIM \cite{mgim} all use simple classification frameworks with predefined total number of categories covering all possibilities for extraction, which are unable to adaptively capture flexible MES. UMIE \cite{umie}, MMUTF \cite{mmutf}, and MQA \cite{mqa} use generative pre-trained models to transform the MEE task into sequence-to-sequence framework, ensuring adaptability to the flexible MES. Unfortunately, their generative frameworks are still not competent for MEE task. The heavy dependence on external detection tools to pre-extract candidate arguments as in UMIE \cite{umie} and MMUTF \cite{mmutf} increases the risk of error propagation for model misjudging or omitting arguments. The lack of an effective guiding framework to prompt the model to perceive flexible event schema as in UMIE \cite{umie} and MQA \cite{mqa} makes it difficult for models to capture complex MES following the event schema.

The second challenge is \textbf{the absence of multimodal-aligned training data for effective knowledge transfer to MEE task}. As annotations for an MEE dataset are highly labor-intensive, the commonly-used M2E2 dataset \cite{wase} for MEE task only provides test set, while currently available Event Extraction (EE) training sets ACE2005 \cite{ace} and SWiG \cite{swig} are unimodal, containing text-only and image-only EE annotations with distinct schema definitions. Current works \cite{wase,unicl,camel,umie,mmutf,mgim} attempt to transfer knowledge learned from unimodal EE to multimodal by adapting an unimodal weakly supervised training framework. However, each training sample only contains unimodal supervised signals from ACE2005 \cite{ace} or SWiG \cite{swig}, causing model's inability to capture the interdependencies between two modalities contributing to the complete MES. MQA \cite{mqa} tries to circumvent the data issue by employing advanced Multimodal Large Language Models (MLLMs) \cite{blip2} for zero-shot inference but fails to achieve ideal performance. Therefore, model needs multimodal labeled EE data, which has not yet been constructed in current works, for training to achieve effective knowledge transfer to MEE task.

To address the first challenge, we propose a \textbf{Stepwise Schema-Guided Prompting Framework} (SSGPF) with MLLM as backbone. Our framework utilizes MLLM's powerful innate ability of general instruction-following and multimodal understanding, guiding MLLM to handle the complex and flexible MES by decomposing MEE into stepwise prompting tasks and inserting versatile schema-guided information into each step's prompt template. At the initial step, we insert candidate event types and definitions into prompt to guide MLLM for MED. Then we adaptively select the corresponding argument schema of the extracted event type and design multi-step dynamic prompting for MEAE. During each step, we design prompt containing event type and trigger word extracted in the initial step with one selected role label from the argument schema, requiring MLLM to output arguments from both modalities belonging to this role. During MEAE, we develop a text-bridged grounding technique to further enhance MLLM's ability to autonomously discover potential image arguments rather than relying on external tools' pre-extraction.

To address the second challenge, we design an effective method to construct a \textbf{weakly-aligned multimodal event labeled dataset} without human labor for training using existing unimodal annotated datasets. We first train a crossmodal news retriever based on general image-text matching model \cite{clip} on large-scale unlabeled news image-caption corpora, then for each sentence in ACE2005 \cite{ace} whose event type is included in M2E2 \cite{wase} schema, we use the retriever to match it with the most event-relevant image in SWiG \cite{swig} belonging to the same event type. We transform each weakly-aligned image-text pair containing annotations from respective modality to instruction tuning data format. Based on the constructed dataset, we adopt \textbf{parameter efficient instruction tuning} with Low-Rank Adaption (LoRA) \cite{lora} technique to train MLLM under SSGPF, enabling effective and efficient knowledge transfer of MLLM to MEE task. Our main contributions are as follows:

\begin{itemize}
  \item We propose a stepwise schema-guided prompting framework to guide MLLM to effectively handle complex and flexible multimedia event structure for MEE task. Specifically, we employ commonly-used open-source LLaVA-v1.5 \cite{llava1.5} as MLLM backbone, and we design a text-bridged grounding technique to extract visual arguments.
  \item We construct a weakly-aligned multimodal event labeled instruction tuning dataset without human labor. Leveraging this dataset, we adopt LoRA \cite{lora} technique on LLaVA-v1.5 \cite{llava1.5} under our prompting framework for parameter efficient instruction tuning, achieving effective knowledge transfer of MLLM to MEE task.
  \item Our method achieves state-of-the-art performance on M2E2 benchmark \cite{wase}. Further ablation studies prove the effectiveness of our prompting framework and the constructed tuning dataset. To the best of our knowledge, we are the first to effectively adapt MLLM to MEE task.
\end{itemize}

\begin{figure*}[t]
\centering
\includegraphics[width=1.0\textwidth]{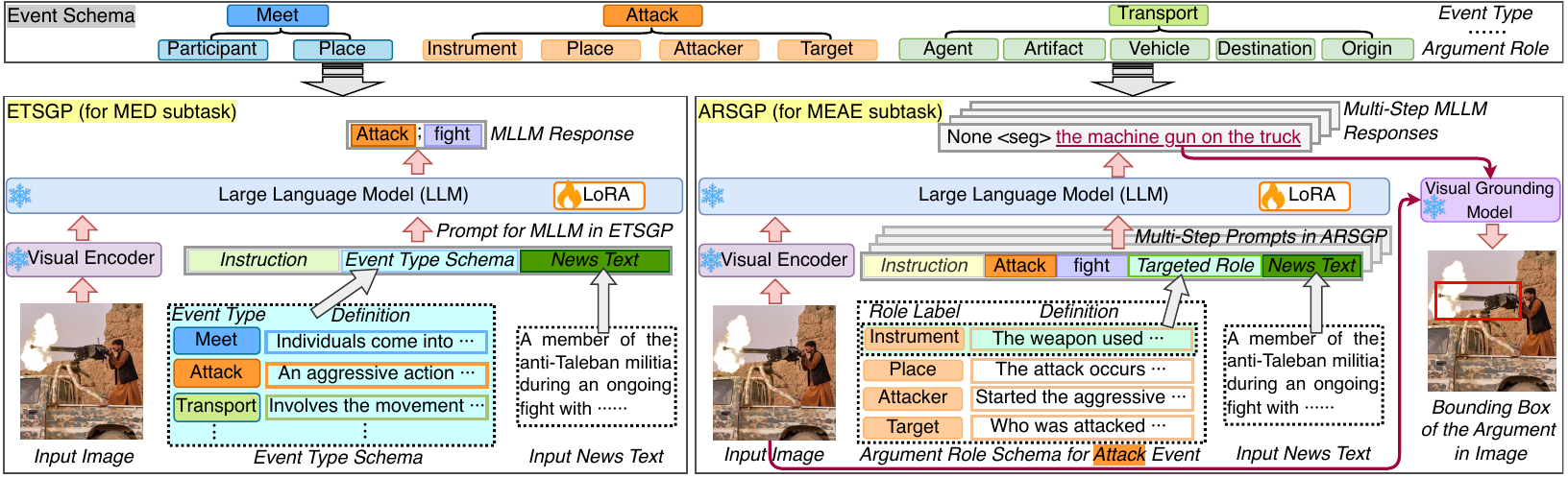}
\caption{Overall workflow of SSGPF. The MLLM backbone consists of a vision encoder and an LLM. At the initial step, we design ETSGP to prompt MLLM to output event type and trigger word. Then we find the argument role schema of the predicted event type in event schema and design ARSGP with multi-step prompting. Each step in ARSGP focuses on one role label, creating adaptive prompt including predicted event type, trigger word, and the targeted role label to prompt MLLM to output all arguments in text and image of this role, where image argument is located by an external visual grounding model according to the generated textual description. During tuning stage, only the injected LoRA modules in LLM of MLLM are trained while all other modules are frozen.}
\label{fig_model}
\vspace{-1em}
\end{figure*}

\section{Related Works}
\subsection{Multimedia Event Extraction} \label{relatedworks}
WASE \cite{wase} is the first to propose MEE task with M2E2 benchmark. It collects unlabeled VOA news image-caption pairs corpora for crossmodal alignment, then conducts unimodal weakly supervised training on graph neural networks for extraction. UniCL \cite{unicl} further uses the VOA corpora with a unified contrastive learning framework for crossmodal gap bridging. CAMEL \cite{camel} uses generative models to conduct crossmodal data augmentation while the generated contents have no annotation. MGIM \cite{mgim} further enhances graph representations in \cite{wase} and conducts multi-grained gradual inference. MMUTF \cite{mmutf} only tackles MEAE, using argument role schema as template query to generate matching vectors with each pre-extracted candidate argument. UMIE \cite{umie} integrates pre-extracted image objects with simple MEE task-related instructions to force LLM to generate MEE outputs, where image arguments are extracted by outputting objects' indices. MQA \cite{mqa} develops a multi-choice question answering framework leveraging MLLM for MED subtask. Despite their efforts, these methods still face challenges introduced in section \ref{sec:intro}. 

\subsection{Large Language Models for Information Extraction Tasks}
The ability of LLMs to effectively handle IE tasks is still under-explored, as IE tasks require models to output knowledge-dense schema-followed structured content. Recent work \cite{llmie} utilizes LLMs' in-context learning ability to prompt LLMs to output structure-organized text following IE schema. It tests representative LLMs on various text-only IE tasks including Named Entity Recognition (NER), Relation Extraction (RE), and Event Extraction (EE), finding that LLMs are not good zero-shot or few-shot extractors. For Multimodal IE, recent works \cite{mllmie1,mllmie2,mqa} explore the zero-shot reasoning ability of representative Multimodal LLMs (MLLMs) for Multimodal NER, Multimodal RE, and subtask MED of MEE and observe that MLLMs perform poorly. These exploratory works analyze that LLMs have limited zero-shot knowledge transfer ability to downstream IE tasks due to the scarcity of IE-related datasets during LLMs' pre-training and instruction-tuning stage and the high flexibility of IE tasks output format. 

\section{Methods}
\subsection{Problem Formulation}
Giving input news image \(I\) and sentence \(T\), the goal of MEE is twofold. (1) MED: extract a set of event mentions, where each event mention \(e\) has an event type \(\xi\) and is grounded on both the given image and a trigger word \(\psi\) in \(T\), model needs to output \((\xi, \psi)\) for \(e\). (2) MEAE: for each spotted event mention \(e\), extract a set of arguments of \(e\), where each argument \(a\) has a role label \(\omega\), and is grounded on an entity \(t\) in sentence \(T\) or an object \(o\) represented as a bounding box in image \(I\), or both. Model needs to output \((\omega,\{t,o\})\) for \(a\), where \(t\) and \(o\) can both exist or only one of them exists. Possible event types and argument role labels are predefined in MEE event schema \(\Omega\) provided by M2E2 benchmark \cite{wase}.

\subsection{Stepwise Schema-Guided Prompting Framework}
Fig. \ref{fig_model} demonstrates our Stepwise Schema-Guided Prompting Framework (SSGPF). At the initial step, we design Event Type Schema Guided Prompting (ETSGP) to guide MLLM to output event type and trigger word for MED. Then we devise Argument Role Schema Guided Prompting (ARSGP) containing multi-step prompts for MEAE. At each step in ARSGP, we insert adaptive schema-guided information prompting MLLM to generate possible arguments for one targeted role.

\textbf{\textit{MLLM Backbone}}. We use commonly-used open-sourced LLaVA-v1.5 \cite{llava1.5} as our MLLM backbone. Given image \(I\) and text \(T\) inputs, LLaVA-v1.5 first uses a visual encoder \(\phi_I(\cdot)\) containing a Vision Transformer and a Multi-Layer Perceptron (MLP) projection layer to embed \(I\) into textual space, then combines \(\phi_I(I)\) with tokenized text input and feeds it into the Vicuna-v1.5 LLM \(f(\cdot)\) to generate natural language response:
\begin{equation}
   Res = f(\phi_I(I), T;\Theta)
\end{equation}
where \(\Theta\) denotes the parameters of LLM.

\textbf{\textit{ETSGP}}. ETSGP is the initial step of SSGPF aiming to tackle MED subtask. We design prompt including all possible event types in MEE schema with detailed definitions, asking MLLM to generate possible event types and text trigger words according to the multimodal input. We add instruction requiring MLLM to first output the event type, followed by the trigger word, separated by a semicolon. If multiple possible events are identified, they are separated by a special token \textit{\textless seg\textgreater}. If no event is detected, we ask MLLM to directly output ``None''. Finally we insert news text into the prompt as complete textual input to MLLM and get the response:
\begin{equation}
   \Xi, \Psi = f(\phi_I(I), Pr_{ED}(T);\Theta_{MED})
\end{equation}
where \(\Xi, \Psi\) denotes the set of each spotted event type \(\xi\) and corresponding trigger word \(\psi\) respectively.

\textbf{\textit{ARSGP}}. ARSGP intends to address MEAE subtask by considering each event type spotted by ETSGP to conduct multi-step inferences. For each event type \(\xi \in \Xi\) parsed by ETSGP output, we fetch the corresponding predefined event argument schema \(\Omega(\xi)\) that contains all possible argument role labels of this event type. Instead of asking MLLM to output arguments for all possible role labels at once, we decompose MEAE into multi-step tasks where each step only focuses on one role label, relieving stress for MLLM handling complex MES while allowing MLLM to fully perceive each argument role's semantic information. For each role label \(\omega \in \Omega(\xi)\), we create prompt with descriptions of current event type, trigger word, and the targeted role label with role definition, requiring MLLM to output all arguments belonging to \(\omega\) in both modalities. We instruct MLLM to first output all arguments in text, then output a special token \textit{\textless seg\textgreater} as a separator, and then output all arguments in image. Multiple arguments in the same modality are separated by semicolons. If there is no argument spotted in text or image, we ask MLLM to output ``None''. Each step in ARSGP can be formulated as:
\begin{equation}
   \Gamma_{T}, \Gamma_{I} = f(\phi_I(I), Pr_{AE}(\xi, \psi, \omega,T);\Theta_{MEAE})
\end{equation}
where \(\Gamma_{T}, \Gamma_{I}\) denotes the set of all spotted arguments of role \(\omega\) in text and image respectively.

\textbf{\textit{Text-Bridged Grounding Technique}}. As MEAE requires model to output bounding box for visual argument while MLLM generates natural language, we design this technique to enable MLLM for image argument extraction. At each step in ARSGP, we prompt MLLM to output the textual description of the argument region in image. For each region description \(\gamma_{I} \in \Gamma_{I}\), we use external Visual Grounding (VG) model \cite{seem} which receives the image and region description to output the bounding box locating such region. It is worth noting that the VG model is only used as a post-processing tool for visualization and MEE metrics evaluation without training, while we guide MLLM itself to develop the ability to discover image arguments manifested by such region description. 

\subsection{Weakly-Aligned Multimodal Event Labeled Dataset}
MEE task lacks a training set and there only exists unimodal event annotated datasets ACE2005 \cite{ace} and SWiG \cite{swig}. ACE2005 contains news-related sentences, while each image in SWiG depicts a more general event beyond news and is represented by an activity verb. The event schema definition of M2E2 MEE benchmark \cite{wase} is a subset of ACE2005 event schema, and for SWiG, previous work \cite{wase} has established the mapping from SWiG event schema to MEE event schema, where 98 of 504 SWiG event types are aligned with 8 event types in M2E2. Thus we make a reasonable assumption: as images in SWiG describe more general events, for each ACE2005 sentence whose event type is in MEE schema, there should always exist an image in SWiG of the same event type that is relevant to the event described in the sentence. The sentence and corresponding image can be considered as a weakly-aligned sample describing a multimodal event. 

Following this intuition, we design an effective way to bridge the only-available unimodal ACE2005 and SWiG to construct a weakly-aligned multimodal event annotated dataset for further instruction tuning. Following the event type mapping in \cite{wase}, for each sentence in ACE2005 containing event type in MEE event schema, we fetch all image candidates in SWiG with the same event type, then we find the most event-related image to the given sentence using a crossmodal news retriever. The retriever calculates matching score between the sentence and each image candidate and selects the highest one to form a weakly-aligned multimodal event sample. To obtain a more accurate event-related matching score, we use the powerful crossmodal retrieval model CLIP \cite{clip} as the retriever and further train it on large-scale parallel news image-caption pairs that have no event annotation collected by \cite{wase} with the same training objective as CLIP pre-training stage to enhance its ability for better news-related crossmodal matching. 

After obtaining all weakly-aligned multimodal event samples, we use their original annotations to create multimodal event labeled instruction tuning dataset. For ETSGP, we use each sample's event type and trigger word to create gold answers following the output format of ETSGP. For ARSGP, we create gold answers for each argument role following the multi-step prompts, where each gold answer contains entity words in text and region description in image belonging to the current targeted role label. We use an image captioning model \cite{blip2} to generate the region description by sending the cropped image using the argument's bounding box to the model and asking it to generate gold description of the argument.

\subsection{Parameter Efficient Instruction Tuning}
To achieve effective knowledge transfer of MLLM to MEE task, we use LoRA technique \cite{lora} to parameter-efficiently fine-tune the MLLM under SSGPF. We inject LoRA module consisting of trainable rank decomposition matrices into each linear layer of LLM in the MLLM backbone, keeping all other parameters frozen. We set up a separate set of LoRA parameters for ETSGP and ARSGP as they have different output formats aiming to complete MED and MEAE respectively. Using our constructed dataset, we perform instruction tuning on MLLM backbone under ETSGP for MED and under ARSGP for MEAE with their respective instructions and gold answers. We adopt teacher-forcing training where candidate argument roles in prompts of ARSGP come from the ground truth event type's argument schema. After tuning, the trained LoRA module is integrated with original LLM parameters for MEE inference, where the candidate argument roles in prompts of ARSGP come from the predicted event type by ETSGP.

\section{Experiments}
\subsection{Experimental Settings}
\textbf{\textit{Datasets and Evaluation}}. We conduct evaluation on commonly-used M2E2 benchmark \cite{wase}. M2E2 contains 309 news image-text pairs with MEE annotations for evaluation, which includes 8 possible event types inherited from text-only ACE2005 \cite{ace} event schema. Our weakly-aligned multimodal event labeled dataset for instruction tuning contains 2,316 image-text pairs constructed from ACE2005 \cite{ace} and SWiG \cite{swig}. More details of these datasets can be found in supplemental material. Following previous MEE works \cite{wase, unicl, camel, mgim}, we report Precision (\textit{P}), Recall (\textit{R}), and F1 Score (\textit{F1}) on MED and MEAE. Specifically, following \cite{wase, unicl, camel, mgim}, an image argument is considered as correctly localized if the IoU score of the predicted bounding box with the ground truth is over 0.5.

\textbf{\textit{Baselines}}. We compare SSGPF with current State-Of-The-Art (SOTA) MEE methods, including FLAT \cite{wase}, WASE \cite{wase}, UniCL \cite{unicl}, CAMEL \cite{camel}, UMIE \cite{umie}, MGIM \cite{mgim}, and MMUTF \cite{mmutf}. Details of these models have been introduced in section \ref{relatedworks}. FLAT is a baseline version of WASE which replaces graph convolution networks with simple concatenation for crossmodal fusion. WASE\textsubscript{att} and WASE\textsubscript{obj} are two variants of WASE as the former uses attention map for image arguments locating while the latter employs an object detection model. 

\textbf{\textit{Implementations}}. We use the smallest version of open-source LLaVA-v1.5 \cite{llava1.5} pre-trained models \textit{llava-v1.5-7b} as our MLLM backbone. We insert LoRA \cite{lora} module into all linear layers of the LLM in MLLM backbone with rank 128 and alpha 64. ETSGP and ARSGP are trained separately with 15 epochs, batch size 96, learning rate 2e-4, and AdamW optimizer. Supplemental material for more implementation details is available at https://github.com/MartinYuanNJU/SSGPF.

\begin{table}[t]
\caption{Comparison with SOTA methods on M2E2 MEE benchmark. The best result is \textbf{bolded} and the second best is \underline{underlined}.}
\begin{center}
\begin{tabular}{lcccccc}
\toprule

& \multicolumn{3}{c}{\textbf{M2E2 MED}} & \multicolumn{3}{c}{\textbf{M2E2 MEAE}} \\
\cmidrule(r){2-4}\cmidrule(r){5-7}
\textbf{Method} & P & R & F1 & P & R & F1 \\
\midrule
FLAT \cite{wase} & 33.9 & 59.8 & 42.2 & 12.9 & 17.6 & 14.9 \\
WASE\textsubscript{att} \cite{wase} & 38.2 & 67.1 & 49.1 & 18.6 & 21.6 & 19.9 \\
WASE\textsubscript{obj} \cite{wase} & 43.0 & 62.1 & 50.8 & 19.5 & 18.9 & 19.2 \\
UniCL \cite{unicl} & 44.1 & 67.7 & 53.4 & 24.3 & 22.6 & 23.4 \\
CAMEL \cite{camel} & \underline{55.6} & 59.5 & 57.5 & 31.4 & \underline{35.1} & \underline{33.2} \\
UMIE \cite{umie} & - & - & \underline{62.1} & - & - & 24.5 \\
MGIM \cite{mgim} & 46.3 & \underline{69.6} & 55.6 & 25.2 & 21.7 & 24.6 \\
MMUTF \cite{mmutf} & - & - & - & \textbf{39.9} & 20.8 & 27.4 \\
SSGPF(Ours) & \textbf{60.4} & \textbf{72.1} & \textbf{65.7} & \underline{33.8} & \textbf{38.5} & \textbf{36.0} \\
\bottomrule
\end{tabular}
\label{tab_mainresults}
\end{center}
\vspace{-2em}
\end{table}

\subsection{Main Results}
Table \ref{tab_mainresults} presents the performance of our SSGPF and SOTA baselines on M2E2 benchmark. \textit{P}, \textit{R} results on MED and MEAE are not provided in the original paper of UMIE \cite{umie}, and MMUTF \cite{mmutf} can only solve MEAE. The results show that our method significantly outperforms current SOTA baselines. Compared with SOTA method UMIE \cite{umie} using generative framework, we outperform it by 5.8\% \textit{F1} on MED and 46.9\% \textit{F1} on MEAE, indicating the superiority of our framework by decomposing the complete MEE task into stepwise tasks with schema-guided prompts to handle the complex and flexible MES. Compared with SOTA method CAMEL \cite{camel} using classification framework and employing generative models to generate unlabeled images for text-only event annotations and vice versa for data augmentation, we surpass it by 14.3\% \textit{F1} on MED and 8.4\% \textit{F1} on MEAE, proving the usefulness of our constructed wealy-aligned multimodal labeled dataset for effective knowledge transfer to MEE task. We observe our SSGPF achieves the best on all metrics except for \textit{P} on MEAE with the second best compared to MMUTF \cite{mmutf}. As MMUTF depends on external object detection tool to pre-extract a set of image argument candidates, it can achieve relatively high \textit{P} due to the precise bounding box generated by the external tool. However, such dependence severely limits model's flexibility, as the external tool may fail to detect non-general objects specific to news events, such as crowd or military weapons. Thus MMUTF performs poorly on \textit{R}. Rather than relying on pre-extracted tools, we design a text-bridged grounding technique enabling MLLM backbone itself to discover potential image arguments. Despite our \textit{P} on MEAE is 15.3\% slightly lower than MMUTF, our \textit{R} and \textit{F1} significantly outperform it with 85.1\% and 31.4\% improvements, proving the robustness of our method towards complex MES following flexible schema.

\begin{table}[t]
\caption{Ablation studies of prompting framework variants}
\begin{center}
\begin{tabular}{lcccccc}
\toprule
& \multicolumn{3}{c}{\textbf{M2E2 MED}} & \multicolumn{3}{c}{\textbf{M2E2 MEAE}} \\
\cmidrule(r){2-4}\cmidrule(r){5-7}
\textbf{Method} & P & R & F1 & P & R & F1 \\
\midrule
SSGPF & 60.4 & 72.1 & 65.7 & 33.8 & 38.5 & 36.0 \\
SSGPF-zs & 15.6 & 17.5 & 16.5 & 9.1 & 15.6 & 11.5 \\
MQA(-zs) \cite{mqa} & 22.7 & 22.0 & 22.4 & - & - & - \\
SSGPF\textsubscript{jALL} & 57.3 & 68.6 & 62.4 & 8.9 & 7.8 & 8.3 \\
SSGPF\textsubscript{jMEAE} & 60.4 & 72.1 & 65.7 & 25.8 & 24.8 & 25.3 \\
\bottomrule
\end{tabular}
\label{tab_ablation1}
\end{center}
\vspace{-1em}
\end{table}

\begin{figure}[t]
\centering
\includegraphics[width=0.47\textwidth]{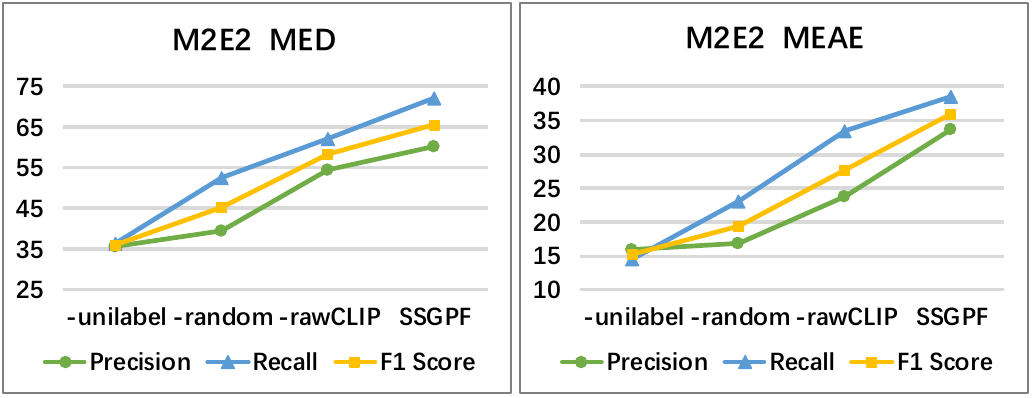}
\caption{Experimental results of training set constructing strategy variants.}
\label{fig_ablation}
\vspace{-1em}
\end{figure}

\subsection{Ablation Studies}
We conduct several ablation studies to further validate the effectiveness of our method. For our prompting framework, we design two variants: (1) SSGPF\textsubscript{jALL}: complete MED and MEAE jointly in one step by designing a single prompt including all candidate event types and argument roles, forcing MLLM to output MEE results in a single response; (2) SSGPF\textsubscript{jMEAE}: retain the initial step ETSGP for MED same as SSGPF, complete MEAE in one step without multi-step prompting by designing a single prompt including all candidate argument roles of the given event type, forcing MLLM to output MEAE results in a single response. We slightly change the instructions' and gold answers' format of our constructed dataset to adapt to these variants for tuning. Moreover, to prove the effectiveness of tuning, we report zero-shot inference results of SSGPF on M2E2 without tuning and a zero-shot reasoning baseline MQA \cite{mqa} which meticulously designs a multiple-choice framework tailored for MED using MLLM while cannot solve MEAE. For a fair comparison, we implement MQA using the same MLLM backbone as SSGPF.

Table \ref{tab_ablation1} presents the results on M2E2 benchmark. ``-zs'' denotes zero-shot results without tuning. SSGPF\textsubscript{jMEAE} maintains ETSGP step the same as full SSGPF, thus their experimental results on MED are identical. We can observe: (1) SSGPF-zs is slightly inferior to MQA, while it significantly outperforms MQA after tuning; (2) Compared with zero-shot inference, SSGPF improves noticeably after tuning; (3) As the prompting framework evolves from SSGPF\textsubscript{jALL}, SSGPF\textsubscript{jMEAE} to SSGPF, performance on MED improves slightly with \(F1\) from 62.4 to 65.7, while on MEAE improves significantly with \(F1\) from 8.3, to 25.3, then to 36.0. These indicate: (1) Even with deliberately designed prompting framework, MLLM cannot achieve ideal zero-shot performance on MEE task, which needs effective instruction tuning; (2) Our method's outstanding performance comes from the effective prompting framework and further tuning, rather than solely relying on the general multimodal zero-shot reasoning ability of MLLM itself; (3) Our complete SSGPF can better guide MLLM to solve MEE task, especially for MEAE, as MEAE mainly reflects the complexity and diversity of MES with arguments distributed across multiple modalities with flexible schema, which is a challenge our SSGPF dedicates to address effectively.

To validate the usefulness of our constructed dataset, we explore the matching strategy of finding the matched image in SWiG \cite{swig} to the sentence in ACE2005 \cite{ace}. We set two variants: (1) ``-random'': for each ACE2005 sentence, randomly select one image in SWiG with the same event type; (2) ``-rawCLIP'': directly use the crossmodal news retriever without further training it on VOA image-text corpora before conducting matching. We additionally set (3) ``-unilabel'': follow previous works' \cite{wase,umie} strategy to directly use unimodal labeled ACE2005 and SWiG for training without crossmodal annotations bridging. We train SSGPF based on the instruction datasets constructed through the above strategies. As shown in Fig. \ref{fig_ablation}, from ``-unilabel'', ``-random'', ``-rawCLIP'' to complete SSGPF, the performance on M2E2 benchmark gradually and significantly increases, which indicates: (1) Model needs to receive multimodal supervised signals rather than unimodal for more effective knowledge transfer; (2) Using the retriever to match image with text can help building higher quality multimodal labeled samples; (3) Further training on VOA corpora enhances the news retriever to find more event-relevant image to the text, improving the constructed dataset's quality.

\subsection{Case Studies}

\begin{figure}[t]
\centering
\includegraphics[width=0.485\textwidth]{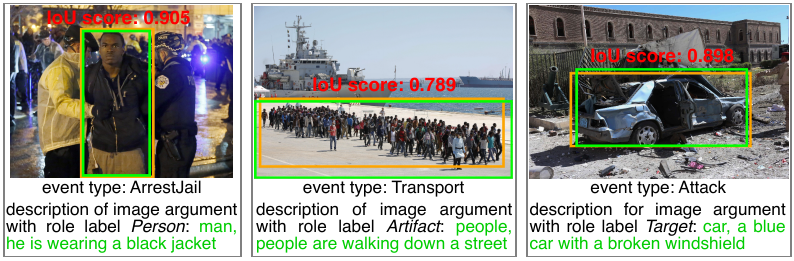}
\caption{Visualizations of SSGPF image argument extraction on M2E2 test set. Text in green color is the predicted description for the argument in image of the given targeted role label. Green bounding box in image denotes the argument's visual region outputted by the visual grounding model based on the predicted description. Orange bounding box in image is the ground truth.}
\label{fig_case}
\vspace{-1em}
\end{figure}

We provide case studies to visualize the text-bridged grounding technique in SSGPF for image argument extraction. In Fig. \ref{fig_case}, we provide three image samples from M2E2 test set to demonstrate the visual input and SSGPF's output of event type for MED and the output of one specified argument role for MEAE. As shown in Fig. \ref{fig_case}, the predicted bounding box in each example is very close to the ground truth, with the IoU score significantly higher than 0.5. This is because the textual description of the predicted image argument generated by MLLM backbone under SSGPF captures the prominent feature of the target region, allowing the VG model to localize the visual area smoothly. The visualization results further confirm the effectiveness of our SSGPF for handling MEE task.

\section{Conclusion}
This work proposes a \textbf{S}tepwise \textbf{S}chema-\textbf{G}uided \textbf{P}rompting \textbf{F}ramework with Parameter Efficient Instruction Tuning (SSGPF) for Multimedia Event Extraction (MEE) task. Our framework employs Multimodal Large Language Model (MLLM) as backbone, decomposing MEE into stepwise prompting tasks with adaptable schema-guided prompt at each step, enabling MLLM to handle the complex and flexible multimedia event structure. As there lacks an annotated MEE training set, we construct a weakly-aligned multimodal labeled MEE instruction tuning dataset from the only-available unimodal event labeled datasets through crossmodal annotations bridging. Then we conduct parameter efficient instruction tuning with LoRA to achieve effective knowledge transfer of MLLM to MEE task. Comprehensive experiments on the M2E2 benchmark demonstrate the effectiveness and superiority of our approach.

\bibliographystyle{IEEEtran}
\bibliography{ref}

\begin{thebibliography}{10}
\providecommand{\url}[1]{#1}
\csname url@samestyle\endcsname
\providecommand{\newblock}{\relax}
\providecommand{\bibinfo}[2]{#2}
\providecommand{\BIBentrySTDinterwordspacing}{\spaceskip=0pt\relax}
\providecommand{\BIBentryALTinterwordstretchfactor}{4}
\providecommand{\BIBentryALTinterwordspacing}{\spaceskip=\fontdimen2\font plus
\BIBentryALTinterwordstretchfactor\fontdimen3\font minus \fontdimen4\font\relax}
\providecommand{\BIBforeignlanguage}[2]{{%
\expandafter\ifx\csname l@#1\endcsname\relax
\typeout{** WARNING: IEEEtran.bst: No hyphenation pattern has been}%
\typeout{** loaded for the language `#1'. Using the pattern for}%
\typeout{** the default language instead.}%
\else
\language=\csname l@#1\endcsname
\fi
#2}}
\providecommand{\BIBdecl}{\relax}
\BIBdecl

\bibitem{wase}
M.~Li, A.~Zareian, Q.~Zeng, S.~Whitehead, D.~Lu, H.~Ji, and S.-F. Chang, ``Cross-media structured common space for multimedia event extraction,'' in \emph{ACL}, 2020, pp. 2557--2568.

\bibitem{unicl}
J.~Liu, Y.~Chen, and J.~Xu, ``Multimedia event extraction from news with a unified contrastive learning framework,'' in \emph{ACM MM}, 2022, pp. 1945--1953.

\bibitem{camel}
Z.~Du, Y.~Li, X.~Guo, Y.~Sun, and B.~Li, ``Training multimedia event extraction with generated images and captions,'' in \emph{ACM MM}, 2023, pp. 5504--5513.

\bibitem{mgim}
Y.~Liu, F.~Liu, L.~Jiao, Q.~Bao, L.~Sun, S.~Li, L.~Li, and X.~Liu, ``Multi-grained gradual inference model for multimedia event extraction,'' \emph{IEEE TCSVT}, vol.~34, no.~10, pp. 10\,507--10\,520, 2024.

\bibitem{umie}
L.~Sun, K.~Zhang, Q.~Li, and R.~Lou, ``Umie: Unified multimodal information extraction with instruction tuning,'' in \emph{AAAI}, 2024, pp. 19\,062--19\,070.

\bibitem{mmutf}
P.~Seeberger, D.~Wagner, and K.~Riedhammer, ``{MMUTF}: Multimodal multimedia event argument extraction with unified template filling,'' in \emph{Findings of EMNLP}, 2024, pp. 6539--6548.

\bibitem{mqa}
Y.~Sun, K.~Zhang, and Y.~Su, ``Multimodal question answering for unified information extraction,'' \emph{arXiv preprint arXiv:2310.03017}, 2023.

\bibitem{ace}
W.~Christopher, S.~Stephanie, M.~Julie, and M.~Kazuaki, ``Ace 2005 multilingual training corpus,'' in \emph{Linguistic Data Consortium, Philadelphia}, 2006.

\bibitem{swig}
S.~Pratt, M.~Yatskar, L.~Weihs, A.~Farhadi, and A.~Kembhavi, ``Grounded situation recognition,'' in \emph{ECCV}, 2020, pp. 314--332.

\bibitem{blip2}
J.~Li, D.~Li, S.~Savarese, and S.~Hoi, ``Blip-2: Bootstrapping language-image pre-training with frozen image encoders and large language models,'' in \emph{ICML}, 2023, pp. 19\,730--19\,742.

\bibitem{clip}
A.~Radford, J.~W. Kim, C.~Hallacy, A.~Ramesh, G.~Goh, S.~Agarwal, G.~Sastry, A.~Askell, P.~Mishkin, J.~Clark \emph{et~al.}, ``Learning transferable visual models from natural language supervision,'' in \emph{ICML}, 2021, pp. 8748--8763.

\bibitem{lora}
E.~J. Hu, Y.~Shen, P.~Wallis, Z.~Allen{-}Zhu, Y.~Li, S.~Wang, L.~Wang, and W.~Chen, ``Lora: Low-rank adaptation of large language models,'' in \emph{ICLR}, 2022.

\bibitem{llava1.5}
H.~Liu, C.~Li, Y.~Li, and Y.~J. Lee, ``Improved baselines with visual instruction tuning,'' in \emph{CVPR}, 2024, pp. 26\,296--26\,306.

\bibitem{llmie}
Y.~Ma, Y.~Cao, Y.~Hong, and A.~Sun, ``Large language model is not a good few-shot information extractor, but a good reranker for hard samples!'' in \emph{Findings of EMNLP}, 2023, pp. 10\,572--10\,601.

\bibitem{mllmie1}
F.~Chen and Y.~Feng, ``Chain-of-thought prompt distillation for multimodal named entity and multimodal relation extraction,'' \emph{arXiv preprint arXiv:2306.14122}, 2023.

\bibitem{mllmie2}
X.~Yang, W.~Wu, S.~Feng, M.~Wang, D.~Wang, Y.~Li, Q.~Sun, Y.~Zhang, X.~Fu, and S.~Poria, ``Mm-bigbench: Evaluating multimodal models on multimodal content comprehension tasks,'' \emph{arXiv preprint arXiv:2310.09036}, 2023.

\bibitem{seem}
X.~Zou, J.~Yang, H.~Zhang, F.~Li, L.~Li, J.~Wang, L.~Wang, J.~Gao, and Y.~J. Lee, ``Segment everything everywhere all at once,'' in \emph{NeurIPS}, 2023, pp. 19\,769--19\,782.

\end{thebibliography}

\end{document}